# Contemporary Shrinking of Colombia's Highest Mountains: Pico Simón Bolivar and Pico Cristóbal Colón

Eric Gilbertson[a]*, Kathryn E. Stanchak[b], Scott Hotaling[c]

[a]*Seattle University, Seattle, USA;* [b]*Issaquah, USA; Utah State University, Logan, UT, USA*

email: gilberte@seattleu.edu

# Contemporary Shrinking of Colombia's Highest Mountains: Pico Simón Bolivar and Pico Cristóbal Colón

Pico Cristóbal Colón and Pico Simón Bolivar are the highest peaks in Colombia and were last accurately surveyed in 1939. This historical survey measured Cristóbal Colón, Colombia's recognized highpoint, at 5775m and Simón Bolivar at 5773m. Both peaks have permanent icecaps at their summits. For this study, multiple differential GPS units and an Abney level were used to re-survey each peak with sub-meter vertical accuracy. As of December 2024, the elevation of Simón Bolivar was 5720.42m +/- 0.08m and Cristóbal Colón was 5712.79m +/- 0.87m. These measurements indicate that the ice caps of both peaks have shrunk dramatically since 1939: Simón Bolivar by 53m and of Cristóbal Colón by 62m. Contrary to official recognition, Pico Simón Bolivar is now the highest mountain in Colombia. Increasing local temperatures suggest that the icecap melting of both peaks is due to climate change.

**Introduction**

Colombia has six glacier masses that cover a total area of 33 square km at elevations ranging from 4800m to over 5700m. Since the early 2000s, when detailed measurements began, the extent of Colombian glaciers has been rapidly decreasing, with an average annual surface area loss rate of 3-5% (Ceballos et al 2023). According to the World Meteorological Organization, glaciers in the tropical Andes have lost 30% of their surface area since the 1980s (WMO 2022).

Although the glacial recession in Colombia has been well-documented, few recent measurements have been taken for glaciers at the highest altitudes in the country, in the coastal Santa Marta Mountains. These mountains host the highest peaks in Colombia: Pico Cristóbal Colón (Colón; location 10.838775S, 73.687191W) and Pico Simón Bolivar (Bolivar; location 10.834718S, 73.690453W; Fig. 1). These adjacent summits are exceptionally prominent in the region, rising to more than 5,500 meters, and visible from long distances. In addition, both Colón and Bolivar are ice-capped summits, meaning the highest point on the peak is ice, and these icecaps are the highest-elevation glaciers in Colombia.

Ice-capped peaks around the world are at risk of climate-induced elevation loss as warming temperatures melt their summit ice (Gilbertson et al., in review). Shrinking of ice-capped summits has already been observed in many mountain ranges including the summit of Kebnekaise, the highest peak in Sweden (Holmlund and Holmlund 2018), the summit of Mt Blanc, the highest peak in France and Italy (Berthier et al 2023), and mountains in Washington, USA, including Mount Rainier, the highest peak in the region (Gilbertson et al., in review). Here, we re-surveyed the highest peaks in Colombia to determine if their ice caps have shrunk since they were last measured in 1939. We conclude by placing our results in the context of recent climate trends in the region.

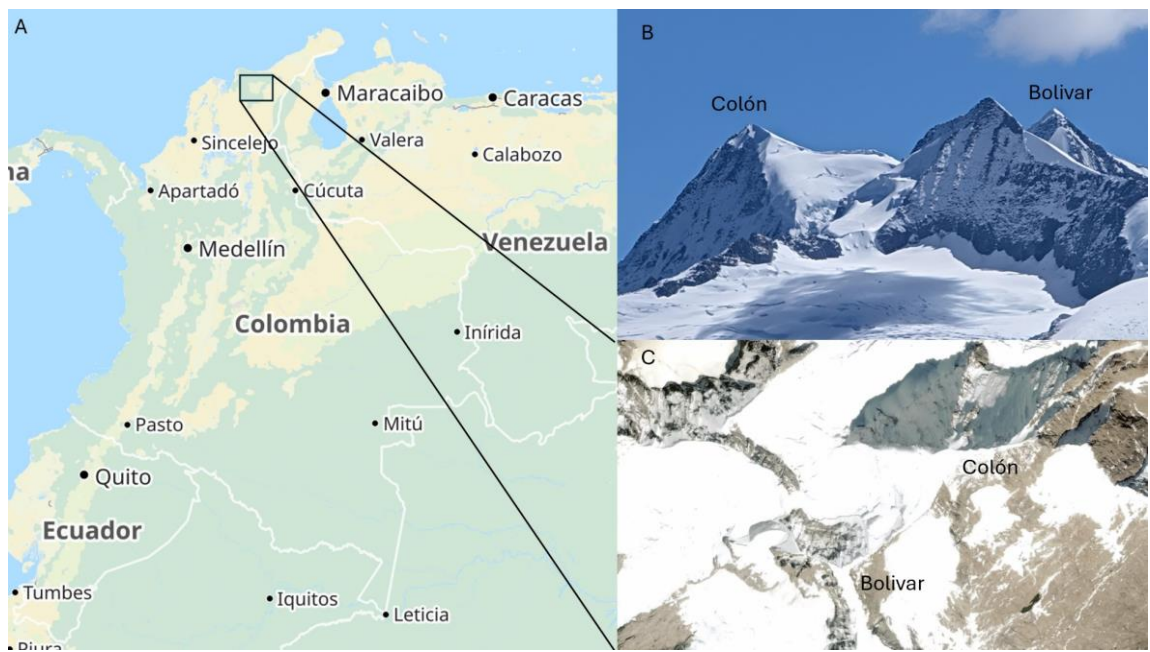

**Figure 1.** A: Location of our focal peaks in the Santa Marta Mountains of Colombia. B: Colón (left) and Bolivar (right), viewed from the north. C: Satellite view of the peaks.

**Methodology**

*History of Climbing and Surveying on Colón and Bolivar*

Bolivar was first climbed on Feb 2, 1939 by Krause, Praolini, and Pichler. Krause surveyed the summit elevation at 5520m using a hypsometer (Ruiz 1990). This measurement was later corrected by the Augustin Codazzi Geographic Institute to 5794m, though the error bounds on the measurement were not given (Ruiz 1990). In general, hypsometers can have errors up to 5-10% of absolute measured altitude (Dobyne, 1988), meaning these measurements could have been in error by approximately 250-500m.

Colón was first climbed a month later, on March 16, 1939, by Praolini, Bakewell, and Wood. This climb was part of the larger American Cabot expedition, which conducted a trigonometric survey of the area. That expedition measured Colón at 5775m and Bolivar at 5773m (Echevarria 1963). Trigonometric surveys in that time period had much lower errors, on the order of a one part per million (0.0001%) (Kershaw 2012). Thus, since 1939, Colón has been recognized as the highest peak in Colombia.

The next survey was conducted in 1989, when Von Rotkirch and Ruiz brought an altimeter to each summit and measured Colón at 5775m and Bolivar at 5790m (Ruiz 1990). In general, altimeters have much higher errors than a trigonometric survey (Dobyne 1988), so the only conclusion that can be reasonably drawn from this survey was that the peaks were of similar elevation in 1989 relative to 1939. Colón was still recognized as the country highpoint.

More recent satellite-based measurements from the 2000 Satellite Range Topography Mission (Farr et al 2007) indicated that the summit of Colón was higher than Bolivar, though vertical errors can be up to 16m for sampled points, even higher for sharp peaks (Sandip 2013), and measurements were only taken every 30m of horizontal spacing (Smith 2023). It is possible, and perhaps likely, that this survey missed the actual highest points on each peak. Thus, the most accurate measurements were still from the 1939 trigonometric survey.

Since the late 1990s, access to this region has been restricted by local indigenous groups. In 2015, Bjorstad and Biggar secured permission to enter the area from the Kogi indigenous community and brought a handheld GPS to the summit of Colón. They measured an elevation of 5725m (GEOCOL2004 geoid). (Bjorstad 2015). In 2022, Alarcon brought a handheld GPS to the summit of Colón and measured 5721m (GEOCOL2004 geoid). Handheld GPS units generally have vertical accuracy around +/- 10m (Rodriguez et al. 2016), so these measurements of Colón are roughly equivalent. The error bounds in these measurements are not known for certain, though. Bolivar was not measured in 2015 or 2022.

*Summit Measurements*

In December 2024, permission was granted by the Arhuaco indigenous community for the lead author (EG) to conduct an elevation survey of both peaks using professional surveying equipment with sub-meter vertical accuracy. To do this, we brought a Trimble DA2 differential GPS (dGPS) unit to the summit of Bolivar on December 23, 2024 at 11am (Fig. 2). GPS data were logged for 45 minutes. In general, in these conditions, sub-meter vertical accuracy is attained after 20 minutes and longer measurements result in improved accuracy. Differential GPS units are capable of 2cm vertical accuracy (Shao and Sui, 2015). Our data were post-processed using two software services: TrimbleRTX (Trimble 2024) and CSRS-PPP (CSRS-PPP 2024). This gave an absolute elevation measurement for Bolivar (denoted as $Z_B$ below).

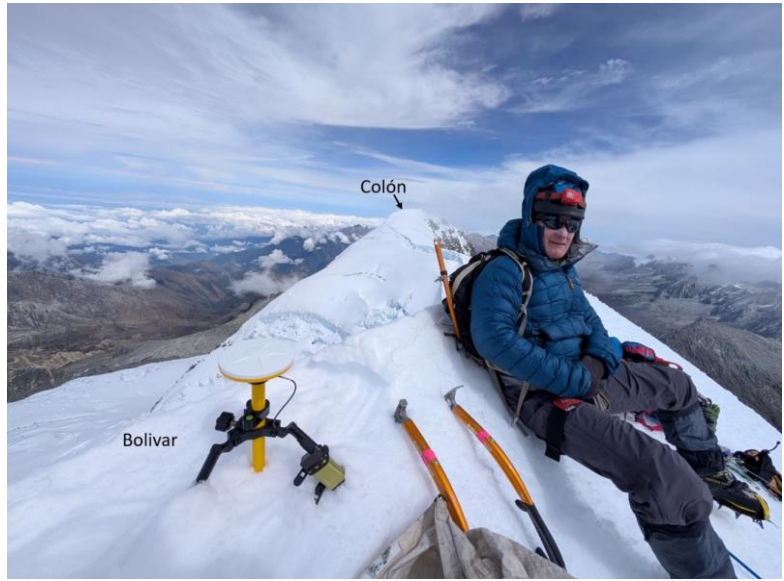

**Figure 2.** The Trimble DA2 differential GPS set up on the summit of Bolivar with bottom of the antenna level with the highest snow. To the east, Colón visible in background.

Given time constraints and the remoteness of the peaks, we were unable to also survey Colón with the dGPS. Instead, we used a standard, highly accurate trigonometric approach to determine the height of Colón from Bolivar (Fig. 3). Specifically, we used a 10 arcminute 5x Sokkia Abney level to measure angular declination from the summit of Bolivar to the summit of Colón. Using known coordinates of each summit, taken from the dGPS measurement and from the 2015 GPS measurement, we then calculated a distance between the summits. Next, we used the declination angle, distance (d), and basic trigonometry to calculate the height of Colón relative to Bolivar (Fig. 3). Finally, we subtracted the relative height of Colón from the measured height of Bolivar to calculate the absolute elevation of Colón, $Z_C$. This is given by:

$$Z_C = Z_B - \Delta \qquad (1)$$

$$\Delta = d \tan \theta \tag{2}$$

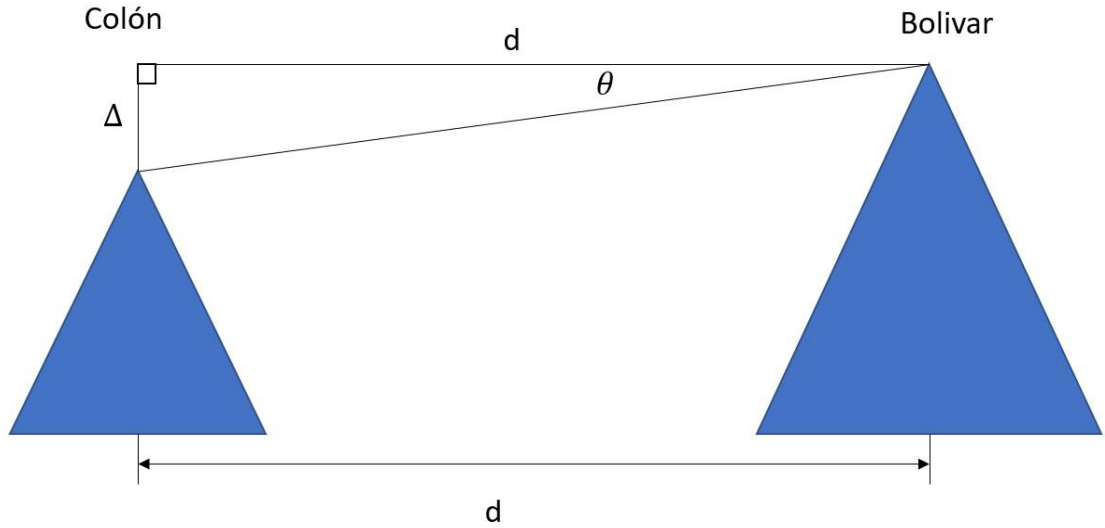

**Figure 3.** Measuring the relative height of Colón from the summit of Bolivar.

*Basecamp Measurements*

To further confirm our summit survey results, we set up two dGPS units (Trimble DA2 and Trimble Promark 220) on a hill near basecamp and recorded data for one hour. We used two units for redundancy of the measurement. In addition, we took Abney level measurements looking up at each peak from the same location (Fig. 4). We then calculated the relative heights of each peak above the dGPS units, using equations:

$$h_1 = d_1 \tan \theta_1 \tag{3}$$

and

$$h_2 = d_2 \tan \theta_2. \tag{4}$$

Distances $d_1$ and $d_2$ were determined using known coordinates of each summit and the measured coordinates of the dGPS units (Fig. 4).

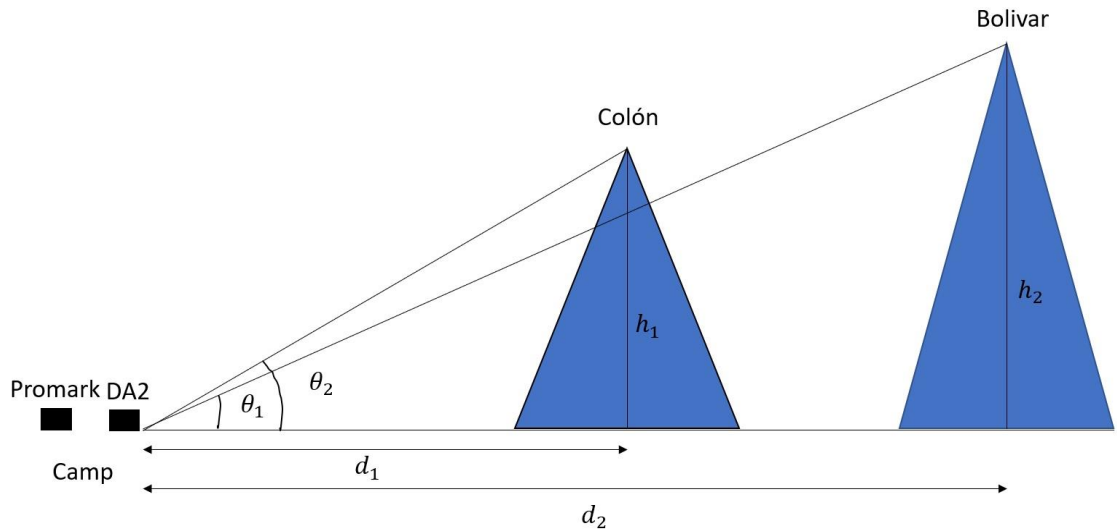

**Figure 4.** Measuring the absolute and relative heights of Colón and Bolivar from basecamp.

Using these measurements, we calculated the relative height between the peaks using the following equation:

$$\Delta = h_2 - h_1 \qquad (5)$$

And, the absolute height of each peak was calculated for Colón ($Z_C$) and Bolivar ($Z_B$) as:

$$Z_C = Z_{camp} + h_1 \qquad (6)$$

$$Z_B = Z_{camp} + h_2 \qquad (7)$$

Where $Z_{camp}$ represents the measured elevation of the dGPS units at camp.

*Approach Pass Measurements*

As a final confirmation of our survey results, we took a dGPS measurement with the DA2 dGPS on a mountain pass during the approach hike that had good visibility to each summit. We paired this with additional measurements of both summits to further validate our findings. To do this, an inclinometer with 0.5-deg resolution and 30x zoom was used to measure the declination $\varphi$ from the summit of Bolivar to the summit of Colon from this location.

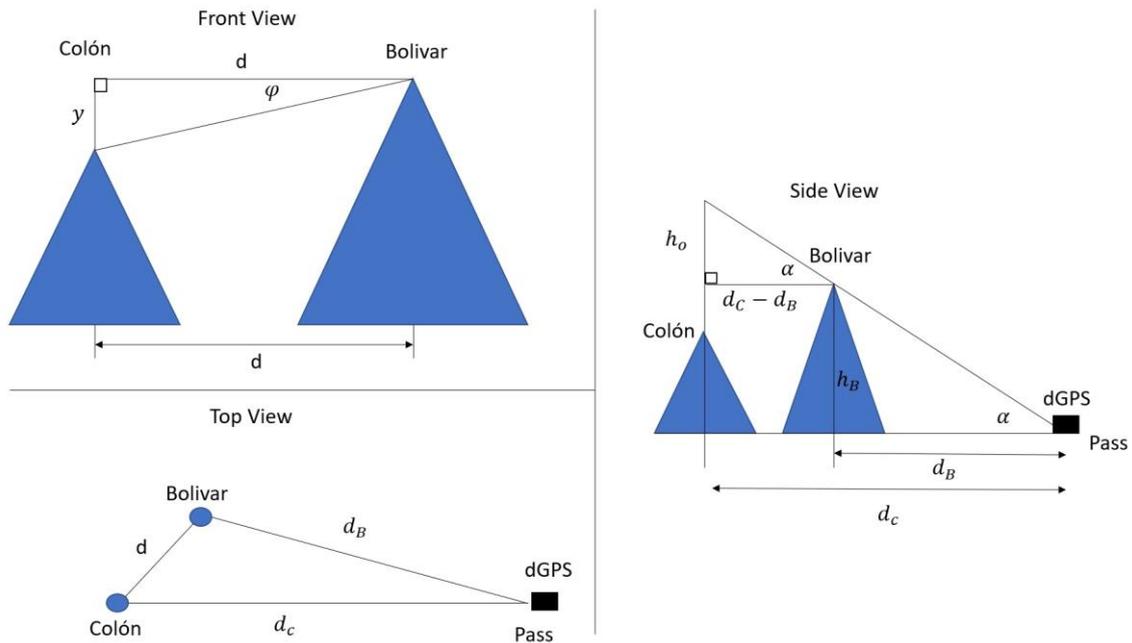

**Figure 5.** Measuring the relative height of Colón and Bolivar from a pass on the approach hike.

If both peaks were equidistant from the pass, then this would mean the relative height would be $y = d \tan \varphi$. However, because Bolivar was closer to the pass than Colón (Fig. 5), this simple equation would overestimate Bolivar's height by $h_o$. Thus, we added a correction to determine the adjusted relative height of Bolivar. The adjusted relative height of Bolivar above Colón was therefore: $\Delta = y - h_o$. This simplifies to:

$$\Delta = d \tan \varphi - (d_c - d_B) \tan \alpha \qquad (8)$$

We found angle $\alpha$ using the right triangle with legs $h_B$ and $d_B$ (Fig 5). This means:

$$\tan \alpha = \frac{h_B}{d_B} = \frac{Z_B - Z_{pass}}{d_B} \quad (9)$$

The distance $d_B$ was measured from the known coordinates of the dGPS and of the summit of Bolivar. The distance $d_C$ was measured from the known coordinates of the dGPS and of the summit of Colón. Thus, the final equation for the relative height between Colón and Bolivar in terms of measured parameters was:

$$\Delta = d \tan \varphi - (d_c - d_B) \frac{Z_B - Z_{pass}}{d_B} \quad (10)$$

From this result, an absolute height of Colón can be calculated by:

$$Z_C = Z_B - \Delta \quad (11)$$

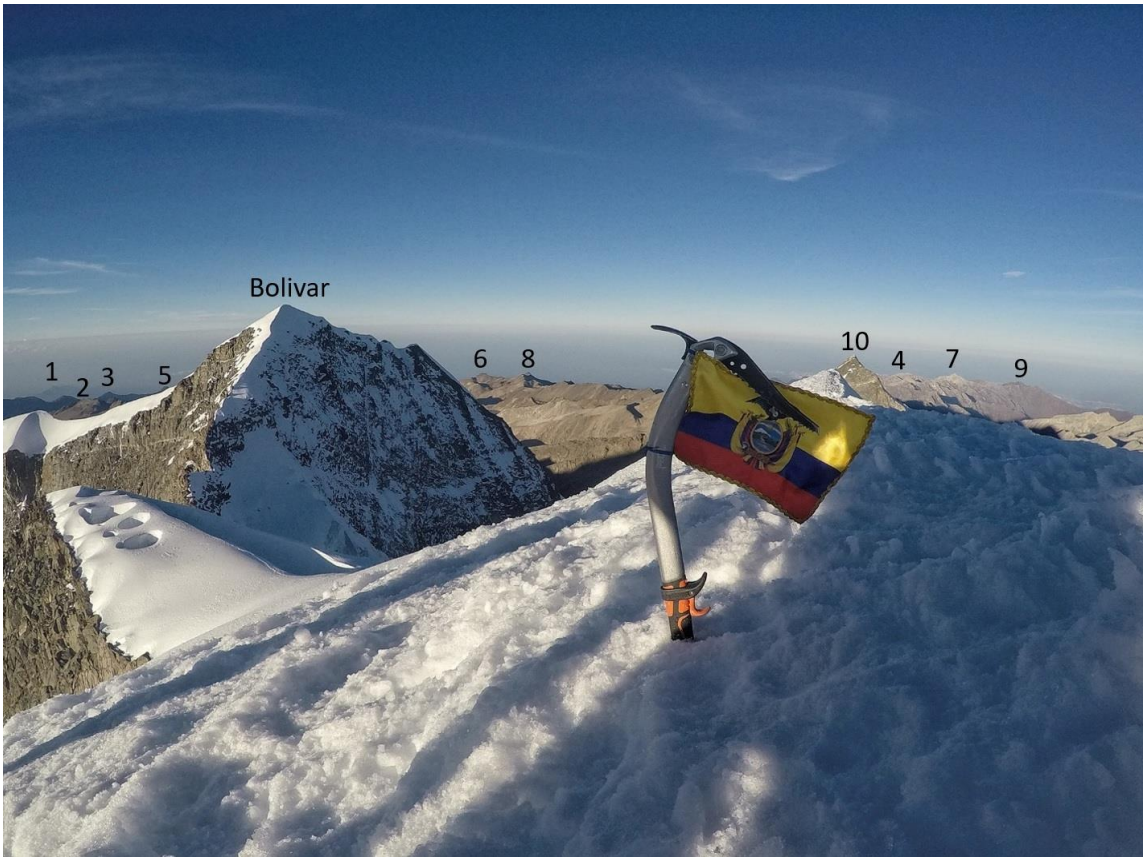

**Figure 6.** A photograph from the summit of Colón looking towards Bolivar from January 2024 used for photographic analysis. Numbers indicate the background peaks described in Table 1.

*Photographic Analysis*

In addition to GPS measurements, we used Geopix photo analysis software (Earl 2017) to analyse a photograph taken in January 2024 from the summit of Colon looking towards Bolivar (Fig. 6). This software allows measurement of the relative height between a peak in the photograph and the camera. It requires entering the coordinates, elevations, and pixel locations of multiple background peaks visible in the photo, as well as the coordinates and pixel location of the peak of interest.

The software accounts for image distortion, atmospheric distortion, and other effects. The software has been validated on over a dozen peaks in Washington, USA, using LiDAR, theodolite surveys, and dGPS surveys. For this analysis, we used ten background peaks (Table 1).

Table 1: Background peaks used for Geopix photographic analysis.

| Peak | Lat | Lon | Elev (m) | Pixel X | Pixel Y |
|---|---|---|---|---|---|
| 1 - Cuchilla de las Coplas | 10.435 | -73.920556 | 1805 | 151 | 713 |
| 2 - Cerro Marmillod South Peak | 10.748889 | -73.744167 | 4530 | 213 | 735 |
| 3 - Pico Achocuimeina | 10.627778 | -73.830833 | 3120 | 266 | 724 |
| 4 - Cuchilla de Yarina | 10.72346 | -73.76607 | 4130 | 270 | 733 |
| 5 - Pico Catorivan | 10.6275 | -73.856111 | 2995 | 386 | 716 |
| 6 - Cuchilla Sigungarua south peak | 10.793611 | -73.783055 | 4660 | 894 | 694 |
| 7 - Peak 4675 | 10.80041 | -73.78078 | 4675 | 947 | 694 |
| 8 - Cuchilla Sigungarua west peak | 10.799722 | -73.785833 | 4680 | 955 | 691 |
| 9 - Peak 4695 | 10.82229 | -73.76835 | 4695 | 1119 | 708 |
| 10 - Simmonds | 10.840809 | -73.705688 | 5620 | 1414 | 653 |

**Results**

*Summit Measurements*

Here and below, we report all elevations as orthometric height using GEOCOL2004 geoid, the reference frame for elevations used in Colombia. For our dGPS summit survey, we measured the summit of Bolivar at 5720.42m +/- 0.08m. Using the Abney level, we measured an angular declination of 40 arcminutes from the summit of Bolivar to the summit of Colón. We used this result to calculate the height of Bolivar relative to Colón to be 7.63m +/- 0.87m. Thus, the absolute height of Colón was 5712.79m +/- 0.87m.

*Additional Measurements and Photographic Analysis*

At basecamp, both dGPS units recorded an elevation of 4561.58 m +/- 0.04m. Using the Abney level, we measured an inclination of 24 degrees 0 arcminutes to Bolivar and 24 degrees 40 arcminutes to Colón. Using the measured coordinates of the dGPS, we determined the horizontal distances to Colón and Bolivar were 2512m and 2608m, respectively. We integrated these results to determine the heights of both peaks at 5714.9m +/- 3.6m for Colón and 5722.8m +/- 3.8m for Bolivar. These measurements assume an error of +/- 5 arcminutes for the Abney level based on its resolution. Thus, these measurements indicate that Bolivar is 7.9m +/- 3.8m taller than Colón.

At the approach pass, we measured an absolute elevation of 4193.60m +/- 0.20m. Our inclinometer measurements indicated an angle of 2.5-3.0 degree declination from the summit of Bolivar to the summit of Colón. Using known and measured coordinates, this meant that Colón and Bolivar were 21.9km and 22.2km from the pass, respectively, and Bolivar's elevation was 1526m above the pass elevation. Using equations (8)-(11), we found Bolivar to be 5.6–10.8m taller than Colón.

According to our photographic analysis, Bolivar is 7.3m +/- 2.4m taller than Colón with a 95% confidence interval.

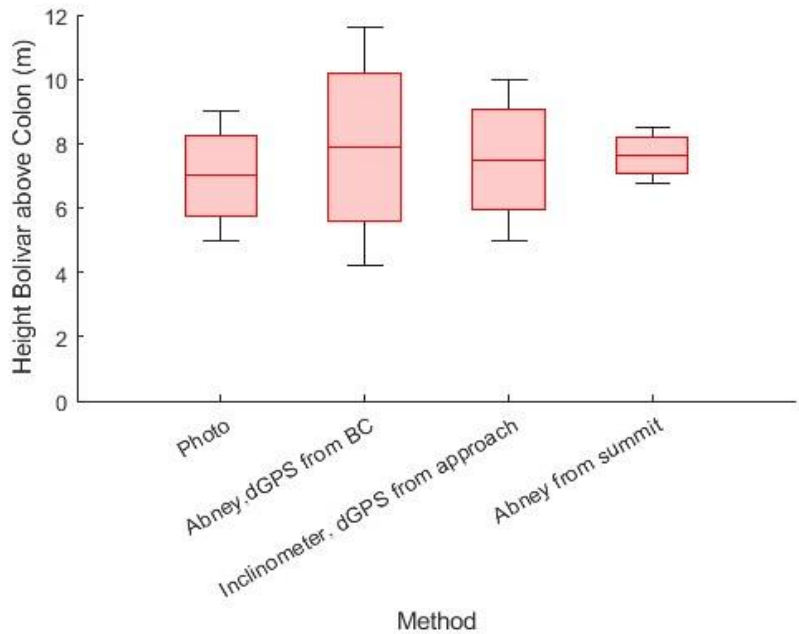

**Figure 7.** Relative heights of Bolivar versus Colón for measurement technique. Boxes are centred around the mean measurement, with the box edges extending to +/- one sigma error, and the whiskers extending to +/- two sigma error, denoting the 95% confidence interval. When sigma values are not known, the whiskers extend to the edges of the error bounds of the measurement.

### *Past and present elevations of Colón and Bolivar*

All four measurement methods are consistent: as of December 2024, Bolivar is taller than Colón (Fig. 7). Integrating the absolute (dGPS) and relative measurements (Abney level), the summit of Bolivar is 5720.42m +/- 0.08m and Colón's summit sits at 5712.79m +/- 0.87m. Bolivar is now 7.63m taller than Colón.

The 2015, 2022, and 2024 measurements of Colón and the 2024 measurement of Bolivar all show significant elevation loss since 1939 (Fig. 8). It appears the elevation loss began in the 1990s or early 2000s. Relative to 1939, the elevations of Bolivar and Colón have declined by 53m and 62m, respectively. This means the icecap on Colón has likely melted at a faster rate than Bolivar.

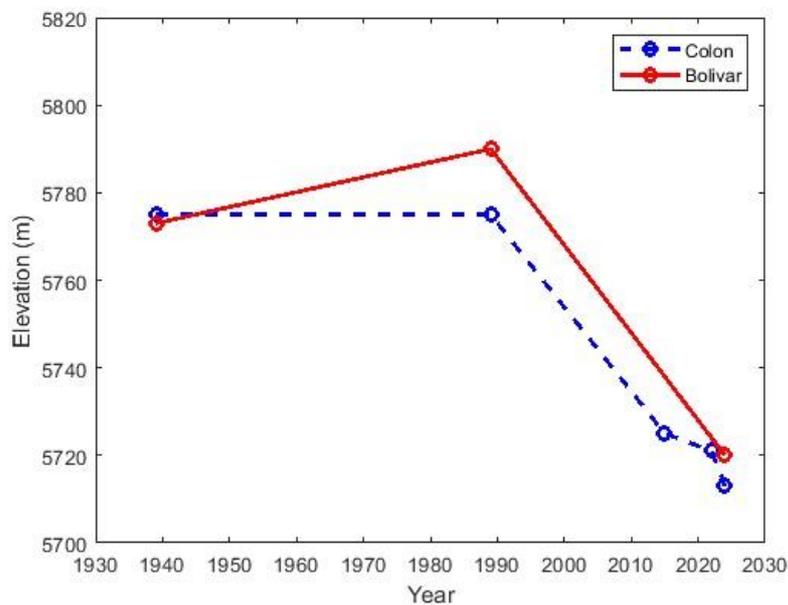

**Figure 8**. Absolute heights of Colón and Bolivar from 1939-2024. Measurements are used are from 1939, 1989, 2015, 2022, and 2024 for Colón and 1939, 1989, and 2024 for Bolivar.

**Discussion**

Pico Simón Bolivar is now the highest mountain in Colombia with an elevation 5720.42m. Pico Cristóbal Colón, formerly the country highpoint in 1939, is now the second highest mountain in Colombia with an elevation of 5712.79m. Both peaks have lost over 50m of elevation since 1939. It is unclear when Bolivar overtook Colón in elevation. Colón was 2m taller than Bolivar in 1939. The 1989 altimeter readings indicated that Bolivar was taller, but the difference was within the error bounds of the measurement. Thus, it is only known that in 1989 Colón and Bolivar were of similar heights to each other, and those heights were similar to the 1939 measurement.

Modelled historical temperature and precipitation anomalies for these summits based on the ERA5 dataset (Hersbach et al 2020) support our theory that climate-induced warming and shifts in precipitation is driving the loss of summit ice for these peaks. Since 2000, 84% of months have had positive temperature anomalies relative to the historical average. And, during the same period, 77% of months have had negative precipitation anomalies. Thus, temperatures are increasing and these warming temperatures are occurring along with precipitation declines. This is likely creating a dual-threat for Colombia's high peaks—warm temperatures are melting summit ice while buffering snow events are also declining. Notably, this dual-threat differs from findings for other ice-capped summits. For instance, in Washington, USA, climate warming is occurring and melting mountain summit ice but this warming is occurring without simultaneous changes in precipitation (Gilbertson et al., in review).

While data remain limited about current and historical elevations of other ice-capped peaks around the world, a clear trend is emerging. We now have evidence from three continents that ice-capped summits are shrinking and climate-induced warming is the most likely driver. These include Europe—Mount Blanc (France/Italy; Berthier et al. 2023), Kebnekaise (Sweden; Holmlund and Holmlund, 2018), North America (Mount Rainier and nearby peaks, USA; Gilbertson et al., in review), and South America (Colón and Bolivar, Colombia; this study). However, the scale of these declines vary; all of the North American and European peaks that have been studied have lost a few to ~25m of elevation since ~1950. At >50m, the scale of elevation decline observed for Bolivar and Colón over a small period far outpaces other mountains. This difference likely stems from two sources. First, the Colombian peaks are closer to the equator and thus more at risk of climate-induced impacts (e.g., Dussaillant et al. 2019). And, second, unlike North America, the Colombian peaks are facing the dual-threat of temperature increase along with precipitation decline.

In this study, we showed that despite their extreme elevation, the highest peaks in Colombia—the first ice-capped peaks to be carefully re-surveyed for elevation changes in South America—are rapidly shrinking. These changes parallel impacts observed in other mountain ranges around the world and highlight the need for repeat surveys of icecaps around the world. Indeed, these types of studies are not just an academic exercise; as is the case in Colombia, they also have implications for cartography and, potentially, tourism. Mountaineers seeking to climb Colombia's highest mountain should shift their focus from Pico Cristóbal Colón to its neighbour, Pico Simón Bolivar.

**Data Availability**

Data is available upon request.

**Acknowledgements**

Surveying equipment was provided by Seattle University and Trimble. Expedition logistics were coordinated by Cristian Alarcon. Permission to survey the peaks was granted by the Arhuaco indigenous community.

**Disclosure statement**

The authors report there are no competing interests to declare.

**Notes on contributors**

Eric Gilbertson is an Associate Teaching Professor in Mechanical Engineering at Seattle University. He has a PhD in Mechanical Engineering from MIT and his research focuses on the effect of climate change on mountain elevations.

Kathryn Stanchak has a PhD from the University of Washington in Biology.

Scott Hotaling is an Assistant Professor in the Department of Watershed Sciences at Utah State University. He leads the Climate Change in Mountains Laboratory (https://qcnr.usu.edu/research/ccml/).